\documentclass[aps,comment,twocolumn,pdftex]{revtex4}
\usepackage{epsfig,amsbsy,bm}
\usepackage{graphicx,color}
\usepackage{dcolumn}
\usepackage{amsmath}
\usepackage{amsfonts}
\usepackage[nice]{nicefrac}
\usepackage[tight]{units}
\usepackage{amssymb,amsmath}
\def\t0{\mbox{$t_{\mbox{{\tiny {0}}}}$}}
\def\p0{\mbox{$p_{\mbox{{\tiny {0}}}}$}}
\def\E0{\mbox{$E_{\mbox{{\tiny {0}}}}$}}

\newcommand{\eref}[1] {(\ref{#1})}

\newcommand{\vect}{\vec}

\renewcommand{\r}{{\bm r}}
  \newcommand{\la}{\langle}
  \newcommand{\ra}{\rangle}
\newcommand{\be}{\begin{equation}}
\newcommand{\ee}{\end{equation}}
\newcommand{\ba}{\begin{eqnarray}}
\newcommand{\ea}{\end{eqnarray}}

\begin{document}
\bibliographystyle{unsrt}
\title
{Photoemission time-delay measurements and calculations close to the 3s ionization minimum in Ar}
\author{D.~Gu\'enot$^1$, K.~Kl\"under$^1$, C.~L.~Arnold$^1$, D.~Kroon$^1$, J.~M.~Dahlstr\"om$^2$, M. Miranda$^1$, T. Fordell$^1$, M.~Gisselbrecht$^1$, P.~Johnsson$^1$, J.~Mauritsson$^1$, E.~Lindroth$^2$, A.~Maquet$^3$, R.~Ta\"ieb$^3$, A.~L'Huillier$^1$ and A.~S.~Kheifets$^4$}

\affiliation{$^1$Department of Physics, Lund University, P.O. Box 118, 22100 Lund, Sweden}
\affiliation{$^2$Department of Physics, Stockholm University, Sweden}
\affiliation{$^3$Laboratoire de Chimie Physique-Mati\`ere et Rayonnement, Universit\'e Pierre et Marie Curie, 11, Rue Pierre et Marie Curie, 75231 Paris Cedex, 05, France}
\affiliation{$^4$Research School of Physical Sciences,
The Australian National University,
Canberra ACT 0200, Australia}
\date{\today}

\begin{abstract}
We present experimental measurements and theoretical calculations of photoionization
time delays from the $3s$ and $3p$ shells in Ar in the photon energy
range of 32-42 eV. The experimental measurements are performed by interferometry
using attosecond pulse trains and the infrared laser used for their
generation.  The theoretical approach includes intershell correlation
effects between the $3s$ and $3p$ shells within the framework of the
random phase approximation with exchange (RPAE).  The connection
between single-photon ionization and the two-color two-photon
ionization process used in the measurement is established using the recently developed asymptotic approximation for the complex transition amplitudes of laser-assisted photoionization. We compare and discuss the theoretical and
experimental results especially in the region where strong intershell correlations in the $3s\to kp$ channel lead to an induced ``Cooper'' minimum in the 3s ionization cross-section.
\end{abstract}

\maketitle

\section{Introduction}

Attosecond pulses created by harmonic generation in gases
\cite{CorkumNaturePhys2007,KrauszRevModPhys2009} allow us to study fundamental light-matter interaction processes in
the time domain.
When an ultrashort light pulse impinges on an atom, a coherent ultra-broadband electron
wave packet is created. If the frequency of the pulse is high enough,
the electronic wave packet escapes by photoionization \cite{SchmidtRPP1992}.
As in ultrafast optics, the {\it group delay}
of an outgoing electron wave packet can be defined by the energy derivative of the phase of the
complex photoionization matrix element.
When photoionization can be reduced to one
non-interacting angular channel ($L$), this phase is the same as the
scattering phase $\eta_L$, which represents the difference between a
free continuum wave and that propagating out of the effective atomic
potential for the $L$-angular channel. In fact, the concept of time
delay was already introduced by Wigner in 1955 to describe $s$-wave
quantum scattering \cite{WignerPR1955}. In collision physics, with
both ingoing and outgoing waves the (Wigner) time delay is twice the
derivative of the scattering phase.

In general, photoionization may involve several strongly interacting
channels. Only in some special cases, the Wigner time delay can be
conveniently used to characterize delay in photoemission. One such
case might be valence shell photoionization of Ne in the 100 eV range
\cite{SchultzeScience2010,KheifetsPRL2010}. In this case, there is no
considerable coupling between the $2s\to  \epsilon p$ and $2p\to
\epsilon s$ or $\epsilon d$ channels and $\epsilon d$ is strongly
dominant over $\epsilon s$, following Fano's propensity rule \cite{FanoPRA1985}.
The case of valence shell photoionization of Ar in the 40 eV range \cite{KlunderPRL2011} is more interesting.
In this case, the $3s$ photoionization is radically modified by strong inter-shell
correlation with $3p$ \cite{Amusia1972361}. As a result, the
$3s$ photoionization cross-section goes through a deep
``Cooper'' minimum at approximately 42 eV photon energy \cite{MobusPRA1993}. Such a feature is a signature of inter-shell
correlation and cannot be theoretically described using any independent electron, e.g. Hartree-Fock (HF) model.

Recent experiments
\cite{SchultzeScience2010,KlunderPRL2011} reported the first
measurements of delays between photoemission from different subshells
from rare gas atoms, thus raising considerable interest from the
scientific community.
Different methods for the measurements of time delays were proposed,
depending on whether single attosecond pulses or attosecond pulse trains were used.
The streaking technique consists in recording electron spectra following ionization
of an atom by a single attosecond pulse in the presence of a relatively intense
infrared (IR) pulse, as a function of the delay between the two pulses \cite{GoulielmakisScience2004,SansoneScience2006}.
Temporal information is obtained by comparing streaking traces from different subshells in
an atom \cite{SchultzeScience2010} or from the conduction and valence bands in a solid \cite{CavalieriNature2007}.
On the other hand, the so-called RABBIT (Reconstruction of attosecond bursts by ionization of two-photon transitions) method consists in recording photoelectron above-threshold-ionization (ATI) spectra following ionization of an atom by a train of attosecond pulses and a weak IR pulse, at different delays between the two fields \cite{PaulScience2001}. Temporal information on photoionization is obtained by comparing RABBIT traces from different subshells in an atom \cite{KlunderPRL2011}. The name of the technique, which we will use throughout, refers to its original use for the measurement of the group delay of attosecond pulses in a train \cite{MairesseScience2003}.

Both methods involve absorption or stimulated emission of one or several IR photons,
and it is important to understand the role of these additional transitions
for a correct interpretation of the measured photoemission delays.
A temporal delay difference of 21 as was measured for the
photoionization from the $2s$ and $2p$ shells in neon using single
attosecond pulses of 100 eV central energy
\cite{SchultzeScience2010}. Interestingly, the electron issued from
$2p$ shell was found to be delayed compared to the more bound $2s$
electron. Similarly, delay differences on the order of $\sim 100$ as
were measured for the photoionization from the $3s$ and $3p$ shells in
argon using attosecond pulse trains with central energy around 35
eV. Again, the $3p$ electron appears to be delayed relative to the
$3s$-electron, with a difference which depends on the excitation
energy \cite{KlunderPRL2011}.

These experimental results stimulated several theoretical
investigations, ranging from advanced photoionization calculations,
including correlations effects \cite{KheifetsPRL2010}, time-dependent
numerical approaches
\cite{SchultzeScience2010,YakovlevPRL2010,NageleJPB2011,TaylorPRA2011} to semi-
analytical developments aiming at understanding the effect of the IR
field on the measured time delays
\cite{ZhangPRA2011,DahlstromPCCP2012,IvanovPRL2011}. The picture which
is emerging from this productive theoretical activity is that
when the influence of the IR laser field is correctly accounted
for, such time delay measurements may provide very interesting
information on temporal aspects of many-electron dynamics.

The present work reports theoretical and experimental investigation of
photoionization in the $3s$ and $3p$ shells in argon in the 32-42 eV photon energy
range. Besides providing a more extensive description of the
experimental and theoretical methods in \cite{KlunderPRL2011}, we
improve the results in three different ways:
\begin{itemize}
\item
We performed more precise measurements using a {\it stabilized} Mach-
Zehnder interferometer \cite{Kroon2012} for the RABBIT method. The
stabilization allows us to take scans during a longer time and thus to
extract the phase more precisely. Some differences with the previous
measurements are found and discussed.
\item
For the comparison with theory, we determined the phases of the single-photon ionization amplitudes
using the Random Phase Approximation with Exchange (RPAE)
method, which includes intershell correlation effects \cite{Amusia1972361,Wendin1972,LHuillierPRA1986}. This represents a clear
improvement to the calculations presented in \cite{KlunderPRL2011},
using Hartree-Fock data \cite{KennedyPRA1972}, especially in the
region above 40 eV where photoionization of Ar passes through an
interference minimum, owing to 3s-3p intershell correlation effects.
\item
Finally, we improved our calculation of the phase of a two-photon
ionization process, thus making a better connection between the
experimental measurements and the single photoionization calculated
phases \cite{DahlstromPCCP2012}.

\end{itemize}

The paper is organized as follows.
Section II presents the experimental setup and
results. Section III describes the phase of one- and two-photon ionization processes using perturbation theory in an independent-electron approximation.
Section IV includes inter-shell correlation using the RPAE method. A
comparison between theory and experiment is presented in
Section V.

\section{Experimental method and results}

\begin{figure}[htbp]
  \centering
  \includegraphics[width=8.0cm]{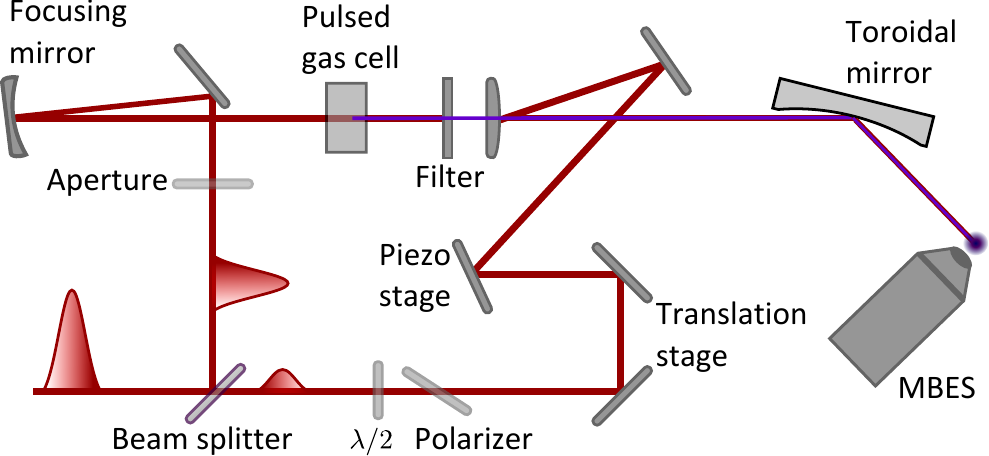}
  \caption{(color online) Schematic illustration of our experimental setup.}
  \label{setup}
\end{figure}

The experiments were performed with a Titanium:Sapphire femtosecond laser system delivering pulses of \(30\)\,fs (FWHM) duration, centered at \(800\)\,nm, with \(1\)\,kHz repetition rate, and a pulse energy of $\sim$ 3 mJ.
A beam splitter divides the laser output into the probe and the pump arm of a Mach-Zehnder interferometer (see Fig. \ref{setup}).
The energy of the probe pulses can be adjusted by a \(\lambda/2\)-plate followed by an ultra-thin polarizer.
The pump arm is focused by a \(\text{f}=50\)\,cm focusing mirror into a pulsed argon gas cell, synchronized with the laser repetition rate, in order to generate an attosecond pulse train via high-order harmonic generation.
An aluminum filter of \(200\)\,nm thickness blocks the fundamental radiation and subsequently a chromium filter of the same thickness selects photon energies of about \(10\)\,eV bandwidth in the range of harmonics 21 to 27.

The probe and the pump arm of the interferometer are recombined on a curved holey mirror, transmitting the pump attosecond pulse train, but reflecting the outer portion of the IR probe beam.
The exact position of the recombination mirror with respect to the focal position of the pump arm is essential in order to precisely match the wavefronts of the probe and XUV (extreme ultraviolet) beams.
A toroidal mirror (\(\text{f}=30\)\,cm) focuses both beams into the sensitive region of a magnetic bottle electron spectrometer (MBES), where a diffusive gas jet provides argon as detection gas.
The relative timing between the ultrashort IR probe pulses and the attosecond pulse train can be reproducibly adjusted on a sub-cycle time scale due to an active stabilization of the pump-probe interferometer length \cite{Kroon2012}.

\begin{figure}[htbp]
  \centering
  \includegraphics[width=8.0cm]{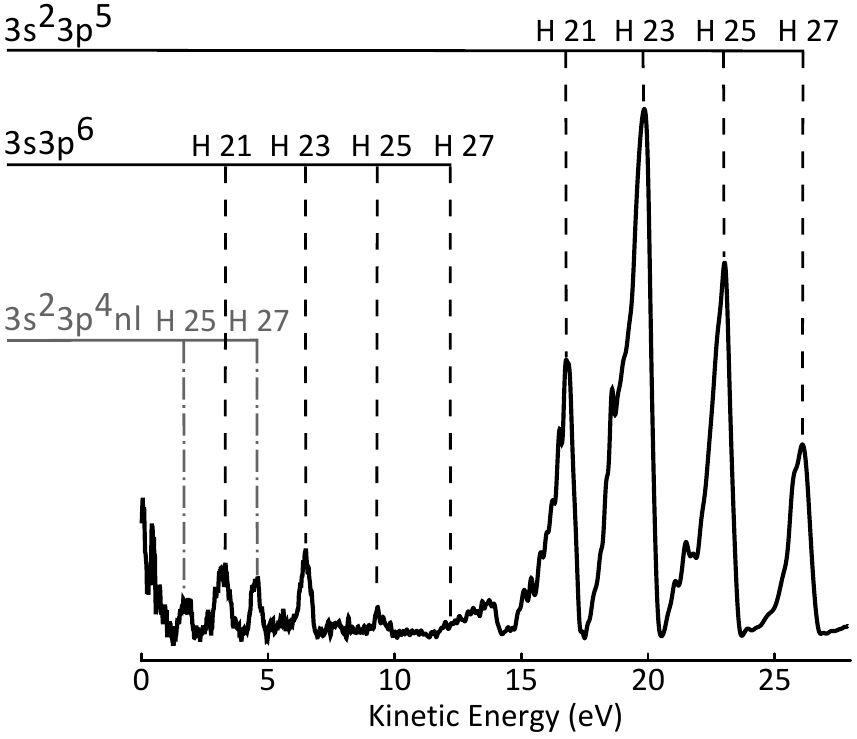}
  \caption{Electron spectrum obtained by ionizing Ar with four harmonics of orders 21, 23, 25 and 27. The ionization channels are shown on the top.}
  \label{Spectroscopy}
\end{figure}

Figure ~\ref{Spectroscopy} presents an electron spectrum obtained by
ionizing Ar atoms with harmonics selected by both Al and Cr
filters, with orders ranging from 21 to 27. We can clearly identify three ionization channels
towards the $3s^2p^5$, $3s^1p^6$, and $3s^23p^4n\ell$ ($n\ell=4p$ or $3d$) continua \cite{KickasJES1996}.
The corresponding ionization energies are 15.76, 29.2 and $\sim$ 37.2 eV. Note that the settings of the MBES
were here chosen to optimize the spectral resolution at low energy. The large asymmetric profile obtained at
high electron energy can be reduced by optimizing
the MBES settings differently. The spectrum due to $3p$-ionization is strongly affected by the behavior of the ionization cross-section in this region. The relative intensities of the 21$^{\textrm{st}}$ to the 27$^{\textrm{th}}$ harmonics are approximately 0.2:0.7:1:1.

\begin{figure}[htbp]
  \centering
  \includegraphics[width=9.0cm]{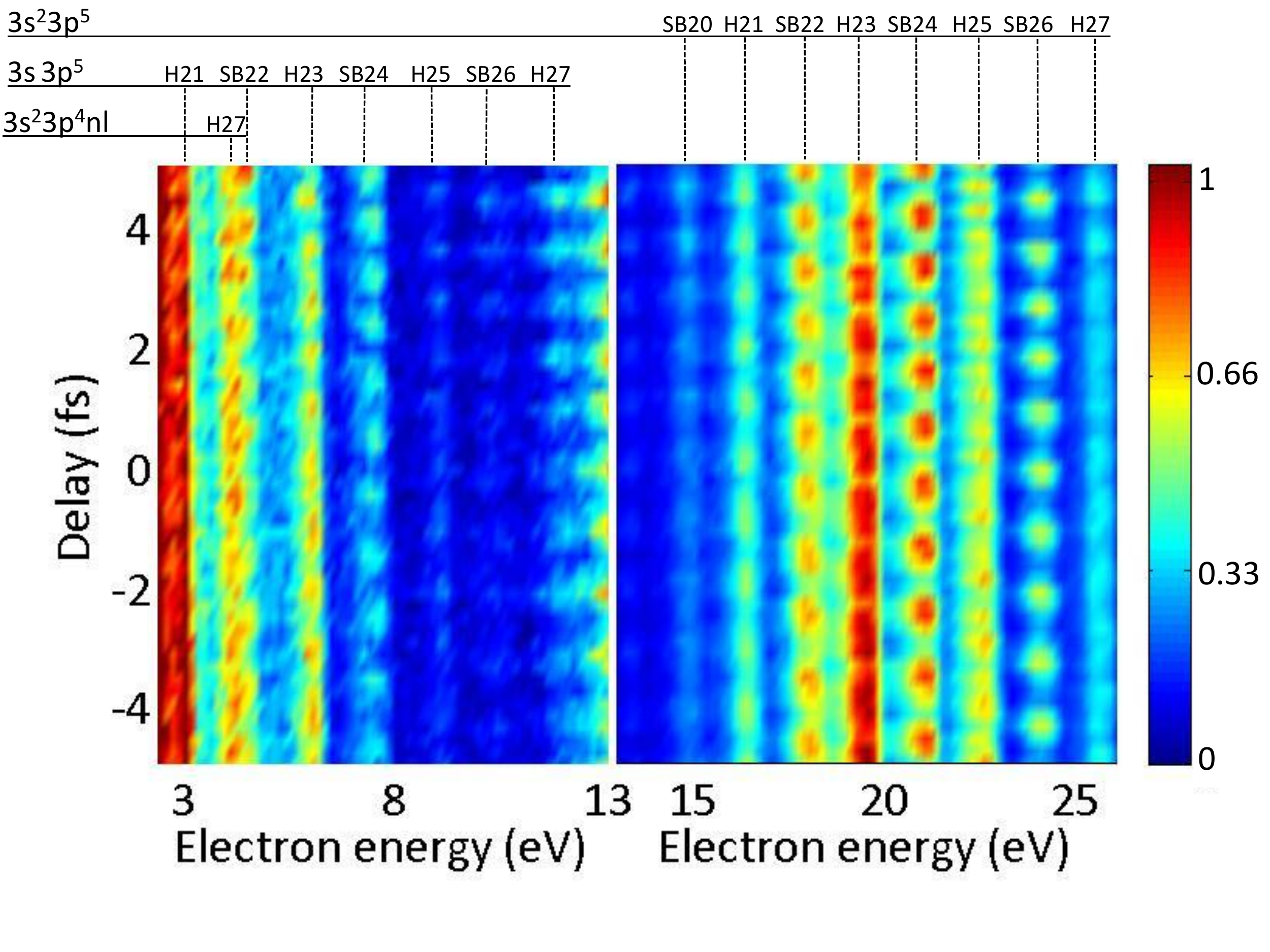}
  \caption{(color online) Electron spectrum as a function of time delay between the attosecond pulses and the IR laser. The signal strength is indicated by colors. The spectrum on the right ($3p$) follows that on the left ($3s$) with a factor of 6 reduction in the color code, and a slight overlap in energy.}
  \label{RABBIT}
\end{figure}

\begin{figure}[htbp]
  \centering
 \includegraphics[width=8.0cm]{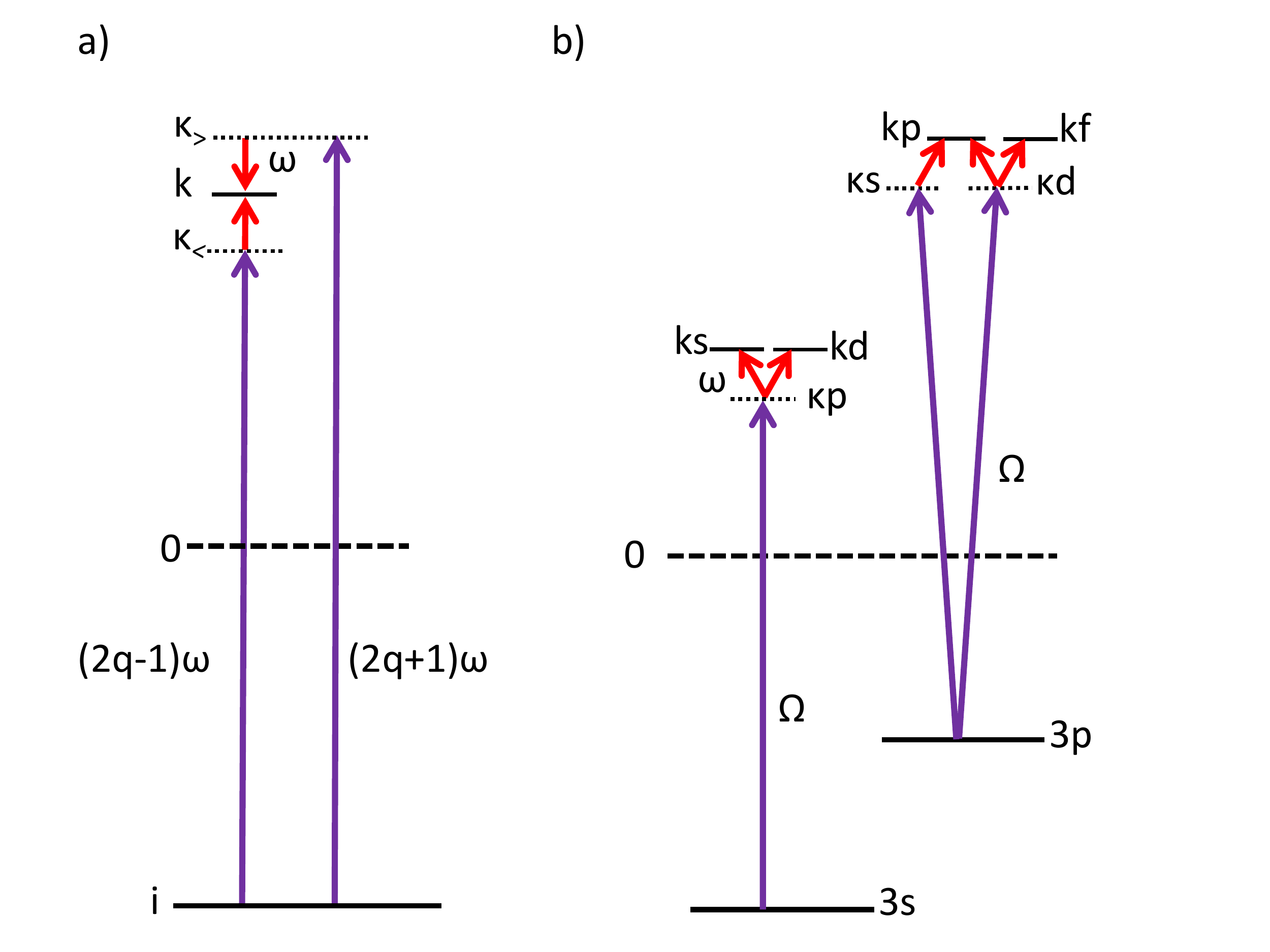}
  \caption{(color online) Energy level scheme of the processes discussed in the present work; (a) RABBIT principle; (b) Different channels in 2-photon ionization from the $3s$ and $3p$ subshells.}
  \label{scheme}
\end{figure}

Figure~\ref{RABBIT} shows a typical RABBIT spectrogram, i.e. electron spectra as a function of delay between pump and probe pulses. The electron yield is indicated as colors. Compared to the spectra obtained with the harmonics only, Fig.~\ref{RABBIT} includes electron peaks at sideband frequencies, including additional absorption or emission of one IR photon (see Fig. \ref{scheme}). The intensity of these sidebands oscillates with a delay at a frequency equal to $2\omega$, $\omega$ being the IR laser photon energy, according to
\be
S_{2q}(\tau)=\alpha + \beta \cos(2\omega \tau-\Delta\phi_{2q}-\Delta\theta_{2q})
\ee
where $\alpha$ and $\beta$ are constant quantities, independent of the delay and $2q$ represents the total number of IR photons involved, i.e. an odd number to create harmonic $2q-1$ or $2q-1$ plus or minus one IR photon. $\Delta\phi_{2q}$ denotes the phase difference between two harmonics with order $2q+1$ and $2q-1$, while $\Delta\theta_{2q}$ arises from the difference in phase between the amplitudes of the two interfering quantum paths leading the same final state [Fig.~\ref{scheme} (a)].
$\tau_A=\Delta\phi_{2q}/2\omega$ can be interpreted as the group delay of the attosecond pulses \cite{MairesseScience2003}. We define in a similar way $\tau^{(2)}= \Delta\theta_{2q}/2\omega$ arising from the two-photon ionization process. Since the same harmonic comb is used for ionization in the $3s$ and $3p$ shells, the influence of the attosecond group delay can be subtracted and the delay difference $\tau^{(2)}(3s)-\tau^{(2)}(3p)$ can be deduced. The results of these measurements are indicated in Table 1 for sidebands 22, 24 and 26. We also indicate in the same table, previous results from \cite{KlunderPRL2011}. It is quite difficult in such an experiment to estimate the uncertainty of our measurement. The stability of the interferometer is measured to be $\sim 50$ as. The relative uncertainty in comparing the phase offsets of different sideband oscillations is estimated to be of the same magnitude or even slightly better.

\begin{table}[h]
\caption{\label{tb}Time delay measurements}
\begin{tabular}{@{}|c|c|c|c|c|}
 \hline
			Sideband & 22& 24& 26\\ \hline
		Photon energy (eV) & 34.1 & 37.2 & 40.3 \\ \hline
		 $\tau^{(2)}(3s)-\tau^{(2)}(3p)$ & & & \\
		 this work (as) &-80 & -100 & 10  \\  \hline
		 $\tau^{(2)}(3s)-\tau^{(2)}(3p)$, & & &  \\
		 \cite{KlunderPRL2011} (as) &-40 (-90) &-110  &-80  \\ \hline
		 $\tau_{cc}(3s)-\tau_{cc}(3p)$ (as)& -150 & -70 & -40 \\ \hline
		 $\tau^{(1)}(3s)-\tau^{(1)}(3p)$(as) &70 &-30 & 50 \\ \hline
\end{tabular}
\end{table}

Our measurements agree well with those of \cite{KlunderPRL2011} for sideband 24.
For sideband 22, the measurements performed in \cite{KlunderPRL2011} could not resolve the sideband peak from electrons ionized by harmonic 27 towards the continuum $3s^23p^4n\ell$ (see Fig.~\ref{RABBIT}). A new analysis done by considering only the high energy part of the sideband peak leads to the number indicated in parenthesis in the table, which is in good agreement with the present measurement. There is, however, a difference for the delay measured at sideband 26. We will comment on this difference in Section V.

\section{Theory of one and two-photon ionization}

To interpret the results presented above, we relate the one-photon ionization delays to the delays measured in the experiment. Using lowest-order perturbation theory, the transition matrix elements in one and two-photon ionization are
\begin{eqnarray}
M^{(1)}(\vec{k}) &=&-i E_\Omega  \langle \vec{k}|z|i\rangle,\\
M^{(2)}(\vec{k}) &=&-i E_\omega E_\Omega \lim_{ \varepsilon \to  0^+} \int \!\!\!\!\!\!\!\! \sum_\nu {\frac{\langle \vec{k}|z|\nu\rangle\langle \nu|z|i\rangle}{\epsilon_i+\Omega-\epsilon_\nu+i\varepsilon}}.
\label{M1}
\end{eqnarray}
Atomic units are used throughout. We choose the quantization axis ($z$) to be the (common) polarization vector of the two fields. The complex amplitudes of the laser and harmonic fields are denoted by $E_\omega$ and $E_\Omega$, with photon energies $\omega$ and $\Omega$, respectively. The initial state is denoted $|i\ra$ and the final state $|\vec{k}\ra$. The energies of the initial and intermediate states are denoted $\epsilon_i$ and $\epsilon_\nu$, respectively. The sum in $M^{(2)}$ is performed over all possible intermediate states $|\nu\ra$ in the discrete and continuum spectrum. The infinitesimal quantity $\varepsilon$ is added to ensure the correct boundary condition for the ionization process, so that the matrix element involves an outgoing photoelectron. The magnitude of the final momentum is restricted by energy conservation to $\epsilon=k^2/2 = \Omega+\epsilon_i$ for one photon and $\epsilon=k^2/2 = \Omega+ \omega+\epsilon_i$ for two-photon absorption. The two-photon transition matrix element involving emission of a laser photon can be written in the same way, with $\omega$ replaced by $-\omega$ in the energy conservation relation and $E_\omega$ replaced by its conjugate.

The next step consists in separating the angular and radial parts of the wavefunctions. The different angular channels involved are indicated in Fig.~\ref{scheme}(b).
We split the radial and angular dependence in the initial state as
$\la \r|i\ra = Y_{l_im_i}(\hat{r})R_{n_il_i}(r)$
and use the partial wave expansion in the final state
\begin{eqnarray}
\label{partial}
\la \r|\vec{k}\ra = (8\pi)^{\frac{3}{2}}
\displaystyle\sum_{L,M}
i^Le^{-i\eta_L(k)}Y^*_{LM}(\hat{k})Y_{LM}(\hat{r})R_{kL}(r).
\end{eqnarray}
We perform the spherical integration in Eq.~(1) and obtain
\ba
\label{dipole}
M^{(1)}(\vect{k})\propto
\sum_{L=l_i\pm1\atop M=m_i}
e^{i\eta_L(k)}i^{-L}Y_{LM}(\hat k) \nonumber\\
\left(\begin{array}{rrr}
L&1&l_i\\
-M&0&m_i\\
\end{array}\right)
\!
T^{(1)}_L(k),
\ea
where the reduced dipole matrix element is defined as
\be
\label{reduced}
T^{(1)}_L(k)=
\hat L \hat l_i
\left(
\begin{array}{rrr}
L&1&l_i\\
0&0&0\\
\end{array}
\right)\\
\la R_{kL}|r|R_{n_il_i} \ra
\ee
using $3j$-symbols and with the notation $\hat l = \sqrt{2l+1}$. The reduced matrix element
\eref{reduced} is real. When the dipole transition with the
increased momentum $L= l_i+1$ is dominant, which is often the case \cite{FanoPRA1985},
 the phase of the complex dipole matrix
element $M^{(1)}$ is simply equal to
\be
\label{argM1}
\arg[M^{(1)}(k)]=\eta_{L}(k)-L\pi/2.
\ee
(There is also a contribution from the fundamental field which we do not write here, as well as
trivial phases, e.g from the spherical harmonic when $M\ne0$ \cite{DahlstromPCCP2012}).
Similarly, for two-photon ionization,

\ba
\label{M2}
M^{(2)}(\vect{k})
\propto
\sum_{L=\lambda\pm1,\lambda=l_i\pm1\atop M=\mu=m_i}
e^{i\eta_L(k)}i^{-L}Y_{LM}(\hat k)\,\\
\left(\begin{array}{rrr}
L&1&\lambda\\
-M&0&\mu\\
\end{array}\right)
\left(\begin{array}{rrr}
\lambda&1&l_i\\
-\mu&0&m_i\\
\end{array}\right)
T^{(2)}_{L\lambda}(k).
\nonumber
\ea
where
\be
\label{reduced2}
T^{(2)}_{L\lambda}(k)=
\hat L \hat \lambda^2 \hat l_i
\left(
\begin{array}{rrr}
L&1&\lambda\\
0&0&0\\
\end{array}
\right)\\
\left(
\begin{array}{rrr}
\lambda&1&l_i\\
0&0&0\\
\end{array}
\right)\\\la R_{kL}|r|\rho_{\kappa\lambda} \ra.
\ee
Here, we have introduced the radial component of the perturbed wave function,
\be
\label{perturbed2}
\vert \rho_{\kappa\lambda} \ra= \lim_{ \varepsilon \to  0^+} \int \!\!\!\!\!\!\!\! \sum_\nu \frac{ \vert R_{\nu\lambda} \ra \la R_{\nu\lambda} \vert r \vert R_{n_il_i} \rangle}{\epsilon_i + \Omega -\epsilon_{\nu}+i\varepsilon}.
\ee
where the sum is performed over the discrete and continuum spectrum. $\kappa$ denotes the momentum corresponding to absorption of one harmonic photon such that the energy denominator goes to zero ($\kappa^2/2= \epsilon_i + \Omega$). The summation can be decomposed into three terms, the discrete sum over states with negative energy, a Cauchy principal part integral where the pole has been removed (both these terms are real) and a resonant term which is purely imaginary. The important conclusion is that in contrast to the radial one-photon matrix element, the radial two-photon matrix element is a complex
quantity.

To evaluate the phase of this quantity, as explained in more details in \cite{DahlstromPCCP2012}, we approximate $R_{kL}(r)$ and $\rho_{\kappa \lambda}(r)$ by their asymptotic values. We have, for example,
 \begin{eqnarray}
\label{perturbed}
&\rho_{\kappa\lambda}(r) & \approx -\sqrt{\frac{2}{\pi\kappa}} \la R_{\kappa\lambda} \vert r \vert R_{n_il_i} \rangle \nonumber \\
& & \!\!\!\frac{1}{r} \exp\left\{i\left[\kappa r+\frac{\ln(2\kappa r)}{\kappa} + \eta_\lambda(\kappa) - \frac{\pi \lambda}{2}\right]\right\}.
\end{eqnarray}
This allows us to evaluate analytically the integral $\la R_{kL}|r|\rho_{\kappa\lambda} \ra$ in Eq.~(\ref{reduced2}). We obtain
\ba
& \arg\left[T^{(2)}_{L\lambda}(k)\right] & \approx (L-\lambda)\frac{\pi}{2} \nonumber \\
& &+\eta_{\lambda}(\kappa) - \eta_{L}(k)+ \phi_{cc}(k,\kappa),
\ea
where $\phi_{cc}(k,\kappa)$ is the phase associated to a continuum-continuum radiative transition
resulting from the absorption of IR photons in the presence of the
Coulomb potential. It is independent from the characteristics of the
initial atomic state, in particular its angular momentum.
An important consequence is that, when inserting the asymptotic form Eq.~(\ref{perturbed}) in Eq.~(\ref{M2}), the scattering phase $\eta_L$ is canceled out, so that {\it the
total phase will not depend on the angular momentum of the final
state}. In the case of a dominant intermediate channel $\lambda$, the phase of the complex two-photon matrix element $M^{(2)}(k)$ is equal to
\be
\arg[M^{(2)}(k)]=\eta_{\lambda}(\kappa)-\lambda\pi/2+\phi_{cc}(k,\kappa).
\ee
It is equal to the one-photon ionization phase towards the intermediate state with momentum $\kappa$ and angular momentum $\lambda$ plus the additional ``continuum-continuum'' phase. The difference of phase which is measured in the experiment is therefore given by
\be
\Delta\theta_{2q}= \eta_{\lambda}(\kappa_>)-\eta_{\lambda}(\kappa_<)
~+~ \phi_{cc}(k,\kappa_>)-\phi_{cc}(k,\kappa_<).
\label{phaseDiff}
\ee
where $\kappa_>$, $\kappa_<$ are the momenta corresponding to the highest (lowest) continuum state in Fig.~\ref{scheme}(a).
Dividing this formula by $2\omega$, we have
\be
\tau^{(2)}(k)=\tau^{(1)}(k)+\tau_{cc}(k),
\ee
where
\be
\tau^{(1)}(k)= \frac{\eta_{\lambda}(\kappa_>)-\eta_{\lambda}(\kappa_<)}{2\omega},
\label{tW}
\ee
is a finite difference approximation to the Wigner time delay $d\eta_{\lambda}/d\epsilon$ and thus reflects the properties of the electronic wave packet ionized by one-photon absorption into the angular channel $\lambda$. $\tau^{(2)}$ also includes a contribution from the IR field which is independent of the angular momentum,
\be
\tau_{cc}(k)= \frac{\phi_{cc}(k,\kappa_>)-\phi_{cc}(k,\kappa_<)}{2\omega}.
\label{tcc}
\ee

\begin{figure}[htbp]
  \centering
 \includegraphics[width=8.0cm]{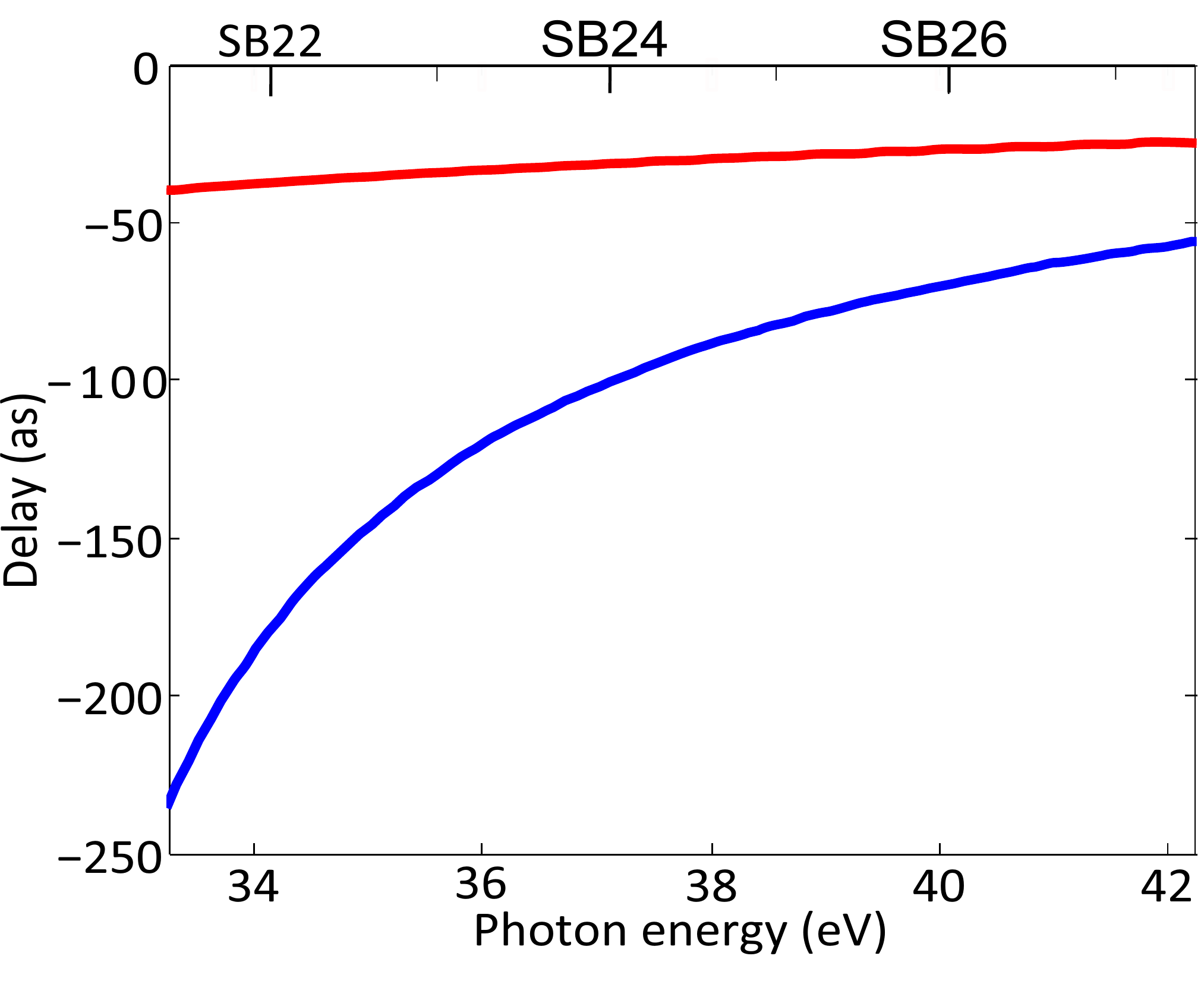}
  \caption{(color online) Continuum-continuum delay $\tau_{cc}$ as a function of excitation photon energy for the two subshells $3s$ (red) and $3p$ (blue) for an IR photon energy of 1.55 eV (800 nm wavelength).}
  \label{continuum}
\end{figure}
We refer the reader to \cite{DahlstromPCCP2012} for details about how to calculate $\tau_{cc}$.
Fig.~\ref{continuum} shows $\tau_{cc}$ as a
function of photon energy, for the two subshells $3s$
and $3p$ and for the IR photon energy
$\omega=1.55$ eV used in the experiment. The corresponding difference in delays for
the $3s$ and $3p$ subshells is only due to the difference in
ionization in energy between the two shells (13.5 eV). We also indicate in Table I the measurement-induced delays
 for the three considered sidebands.

\begin{figure}[htbp]
  \centering
  \includegraphics[width=8.0cm]{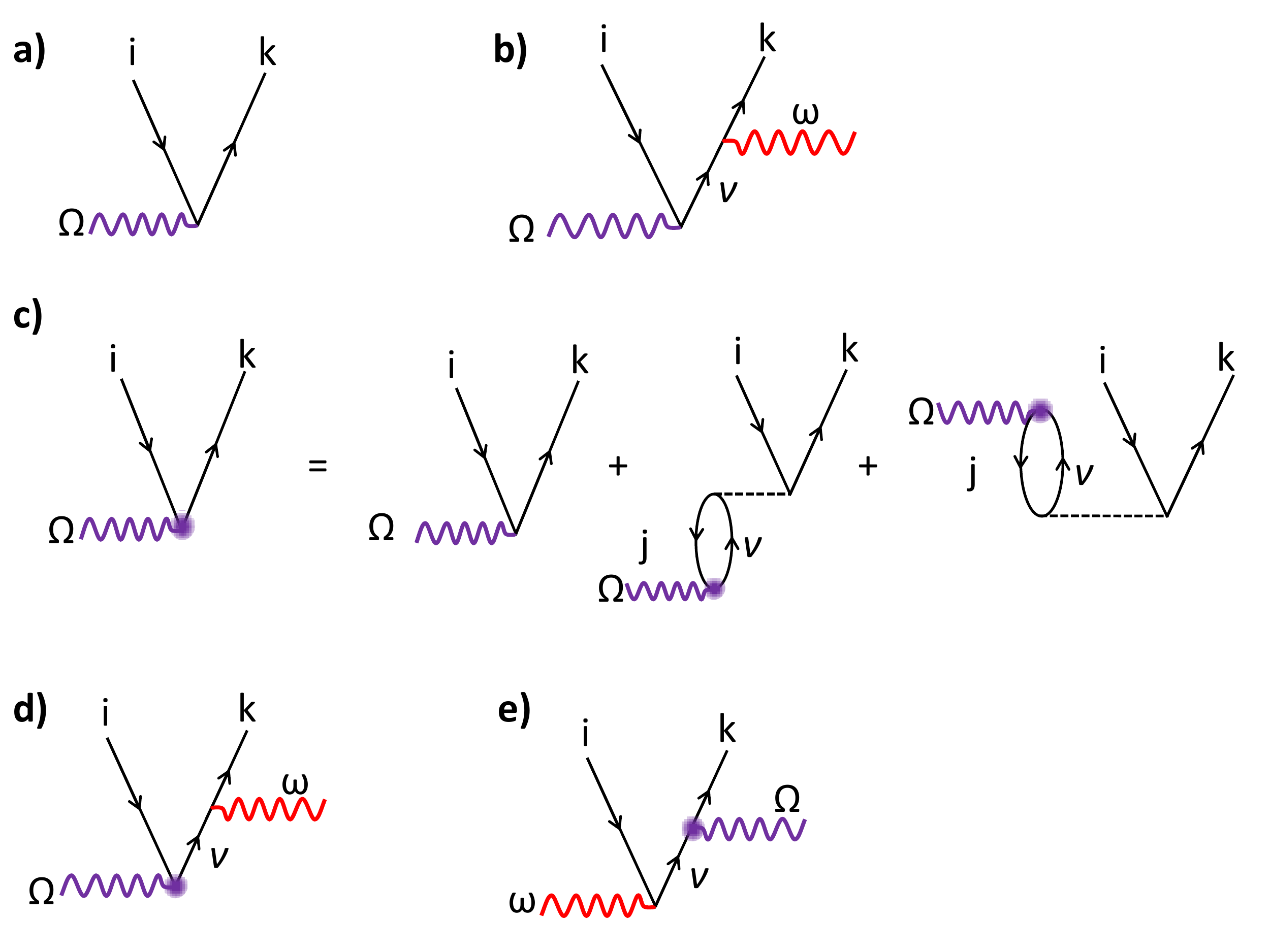}
\bigskip
\caption{(color online) Feynman-Goldstone diagrams representing one-photon (a) and two-photon (b) ionization processes.
(c) Diagrammatic representation of the RPAE equations. The second and third diagrams on the right hand side refer to time-forward and time-reversed respectively. (d) Two-photon ionization including intershell correlation effects. (e) Two-photon ionization with the XUV photon absorbed after the IR photon.}
\label{Feynman}
\end{figure}

The processes discussed in this section can be represented graphically by Feynman-Goldstone diagrams displayed Fig.~\ref{Feynman}(a,b).
The straight lines with arrows represent electron (arrow pointing up) or hole (arrow pointing down) states, respectively.
The violet and red wavy lines represent interaction with the XUV and IR fields. We are neglecting here two-photon processes where the IR photon is absorbed first \cite{DahlstromPCCP2012}.

\section{Intershell correlation effects}

To include inter-shell correlation effects, we use the random phase
approximation with exchange (RPAE) \cite{Amusia1972361}. In this approximation, the
dipole matrix element of single photoionization is replaced by a ``screened'' matrix element $\la k|Z|i \rangle$,
 which accounts for correlation effects between the $3s$ and $3p$ subshells. These screened matrix elements, represented graphically in Fig.~\ref{Feynman}(c), are defined by the self-consistent equation:
\begin{eqnarray}
\label{RPA}
\la \vect k|Z|i \ra &=&
\la \vect k|z|i \ra +
\lim_{ \varepsilon \to  0^+} \int \!\!\!\!\!\!\!\! \sum_{\nu}
\Bigg[ \frac {\la \nu |Z|j \ra \la j \vect k|V|\nu i \ra}
{\Omega - \epsilon_\nu+\epsilon_j +i\varepsilon}
\nonumber\\ &&\hspace*{0.75cm} -
\frac {\la j |Z|\nu   \ra
\la \nu \vect k|V|ji \ra}
{\Omega +  \epsilon_{\nu}-\epsilon_{j}}
\Bigg],
\end{eqnarray}
where $i$ and $j$ are $3s$ or $3p$ or vice versa and $V=1/r_{12}$ is the Coulomb interaction. The sum is performed over the discrete as well as continuum spectrum.
The Coulomb interaction matrices $\la j\vect k|V|\nu i \ra $ and $\la \nu \vect k|V|ij \ra $, represented by dashed lines in Fig.~\ref{Feynman}(c), describe the
so-called time-forward and time-reversed correlation processes (Note that the time goes upwards in the diagrams).
If we replace $Z$ by $z$ in the right term in Eq.~(\ref{RPA}), we obtain a perturbative expansion to the first order in the Coulomb interaction.
More generally, the use of the self-consistent screened matrix elements [Eq.(\ref{RPA})] implies infinite partial sums over two important classes
of so-called ``bubble'' diagrams. Each bubble consists of an
electron-hole pair $\nu j$, which interacts via $1/r_{12}$ with  final
electron-hole pair $\vect ki$.
The energy integration in the time-forward term of Eq.~(\ref{RPA}) (first line) contains
a pole and the screened matrix element acquires an imaginary
part and therefore an extra phase. For a single dominant channel $L$, the phase of the one-photon matrix element
[see Eq.(\ref{argM1})] becomes:
\be
\arg[M^{(1)}(k)]=\eta_{L}(k)+\delta_L(k)-L\pi/2,
\ee
where $\delta_L(k)=\delta_{i\to kL}$ denotes the additional phase due to the correlations accounted within the RPAE.
The photoemission time delay is then
determined by the sum of two terms:
\be
\tau^{(1)} = \frac{d\eta_L}  {d\epsilon}+
\frac{d\delta_L}{d\epsilon}.
\ee
The first term represents the time delay in the independent electron
approximation, equal to the derivative of the photoelectron
scattering phase in the combined field of the nucleus and the
remaining atomic electrons. The second term is the RPAE correction
due to inter-shell correlation effects.

\begin{figure}[htbp]
  \centering
  \includegraphics[width=7.0cm]{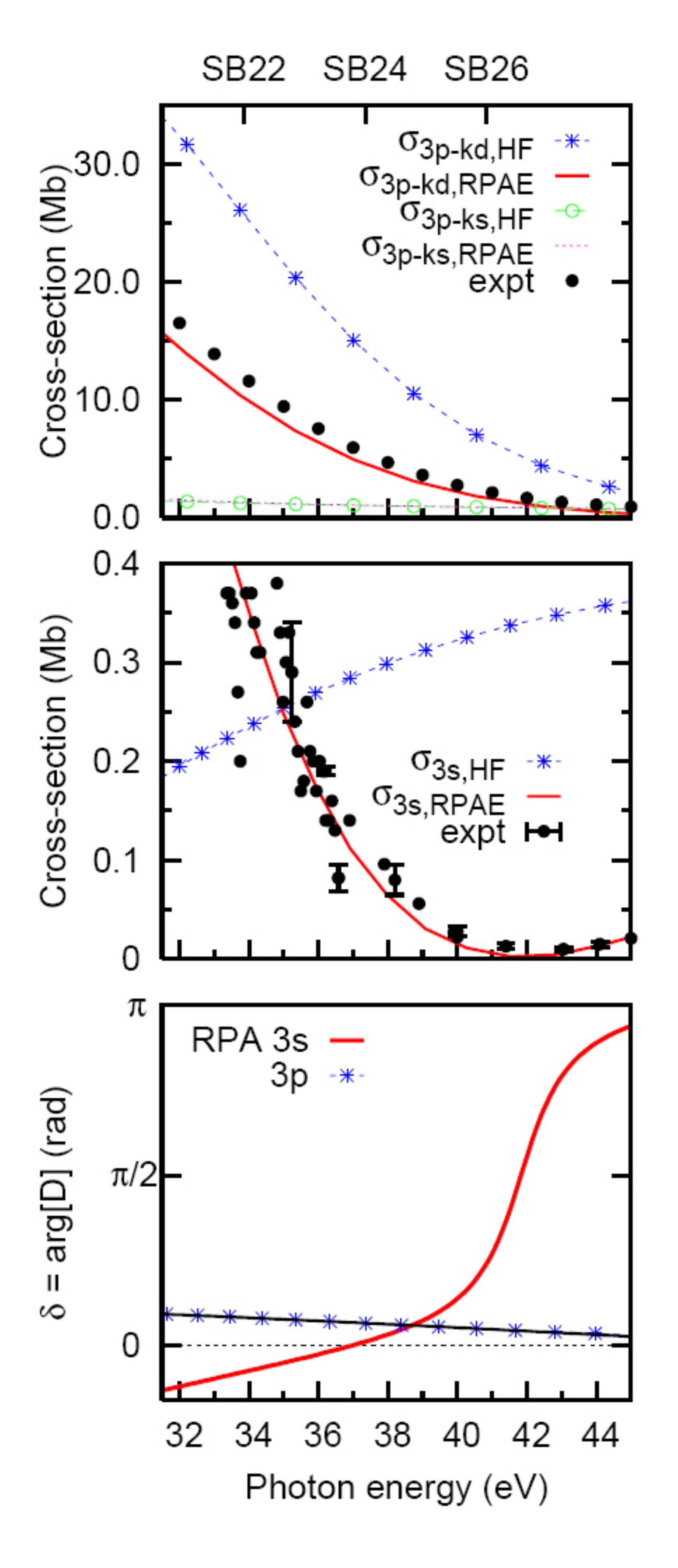}
\caption{(color online) Top panel:
The photoionization cross-sections $\sigma_{3p\to kd}$
calculated in the HF (blue dotted line) and RPAE (red solid line) approximations, are
compared with the $\sigma_{3p\to ks}$ cross-sections
(HF, green open circles, RPAE, purple dashed
line). The experimental data for the $\sigma_{3p}$ cross-section are
from Ref.~\cite{Samson2002265}.
Middle panel: Photoionization cross-sections $\sigma_{3s}$ calculated
in the HF (blue
dotted line) and RPAE(red solid line) approximations, are compared with
experimental data
\cite{MobusPRA1993}.
Bottom panel: Correlation-induced phase
shifts for the $3s$ and $3p$ dipole matrix elements.}
\label{Fig5}
\end{figure}

We solve the system of integral equations \eref{RPA} numerically using
the computer code developed by Amusia and collaborators \cite{AC97}.  The basis of occupied atomic states (holes)
$3s$ and $3p$ is defined by the self-consistent HF method \cite{CCR76}. The excited electron states are
calculated within the frozen-core HF approximation \cite{CCR79}.
We present some results for one-photon ionization in Fig.~\ref{Fig5}.
On the top panel, we show the partial photoionization cross-sections from the $3p$-state calculated using the HF and RPAE approximations (see figure caption).
From this plot, we see that the $3p\to kd$ transition is clearly
dominant at low photon energies. Intershell correlation effects are more important for the $3p\to ks$ than for the $3p\to kd$ transition.
The sum of the two partial cross sections calculated with the RPAE correction (red and green lines) is very close to the
the experimental data (solid circles) \cite{Samson2002265}.
The middle panel presents the calculated cross-section for $3s$-ionization
and compares it to the experimental data from
\cite{MobusPRA1993}. The RPAE correction is here essential to reproduce the behavior of the cross-section which, in this spectral region,
is a rapid decreasing function of photon energy.

The bottom panel shows the
correlation-induced phase shifts
$\delta_{3s\to kp}$ and $\delta_{3p\to kd}$ from the same RPAE calculation.
We observe that the RPAE phase correction $\delta_{3p \to kd}$ is
relatively weak. In contrast, $\delta_{3s\to kp}$ varies
significantly with energy, especially near the ``Cooper''
minimum. This qualitative difference can be explained
by a different nature of the correlations in the $3p$ and $3s$
shells. In the $3p$ case, the correlation takes place mainly between the
electrons that belong to the same shell with not much influence of the
inter-shell correlation with $3s$.
We confirmed this conclusion by performing a
separate set of RPAE calculations with only the $3p$ shell included. These
calculations lead to essentially the same results for $3p$ ionization as the complete calculations.
In the case of intrashell correlation, the
time-forward process [see Fig. \ref{Feynman}(c)] is effectively accounted for by calculating the
photoelectron wave function in the field of a singly charged ion. It
is therefore excluded from Eq.~(\ref{RPA}) to avoid double count. The
remaining time-reversed term [second line in Eq.~(\ref{RPA})] does not contain any poles and therefore does not contribute to an additional phase
to the corresponding dipole matrix element. The small
phase $\delta_{3p\to kd}$ is due to intershell correlation
which is indeed weak. In contrast, $3s$-ionization is strongly affected by correlation with the $3p$ shell.
Consequently, the RPAE phase correction $\delta_{3s\to kp}$, which comes from the correlation with the
$3p$ shell in the time-forward process, is large and exhibits a rapid variation with energy (a $\pi$ phase change) in the region where the cross section decreases significantly.

Finally, we generalize our theoretical derivation of two-photon ionization to including the effect of inter-shell correlation on the XUV photon absorption. As shown graphically in the diagram in Fig.~\ref{Feynman} (d), we replace the (real) transition matrix element corresponding to one-XUV photon absorption by a (complex) screened matrix element, with an additional phase term. As a consequence the phase of the two-photon matrix element becomes:
\be
\arg[M^{(2)}(k)]=\eta_{\lambda}(\kappa) +\delta_{\lambda}(\kappa)-\lambda\pi/2+\phi_{cc}(k,\kappa).
\ee
The time delay measured in the experiment is expressed as before as $\tau^{(2)}(k)=\tau^{(1)}(k)+\tau_{cc}(k)$, with $\tau^{(1)}(k)$ modified by intershell correlation:
\be
\tau^{(1)}(k)= \frac{\eta_{\lambda}(\kappa_+)-\eta_{\lambda}(\kappa_-)}{2\omega} + \frac{\delta_{\lambda}(\kappa_+)-\delta_{\lambda}(\kappa_-)}{2\omega}.
\label{tWRPA}
\ee

\begin{figure}[htbp]
  \centering
 \includegraphics[width=8.0cm]{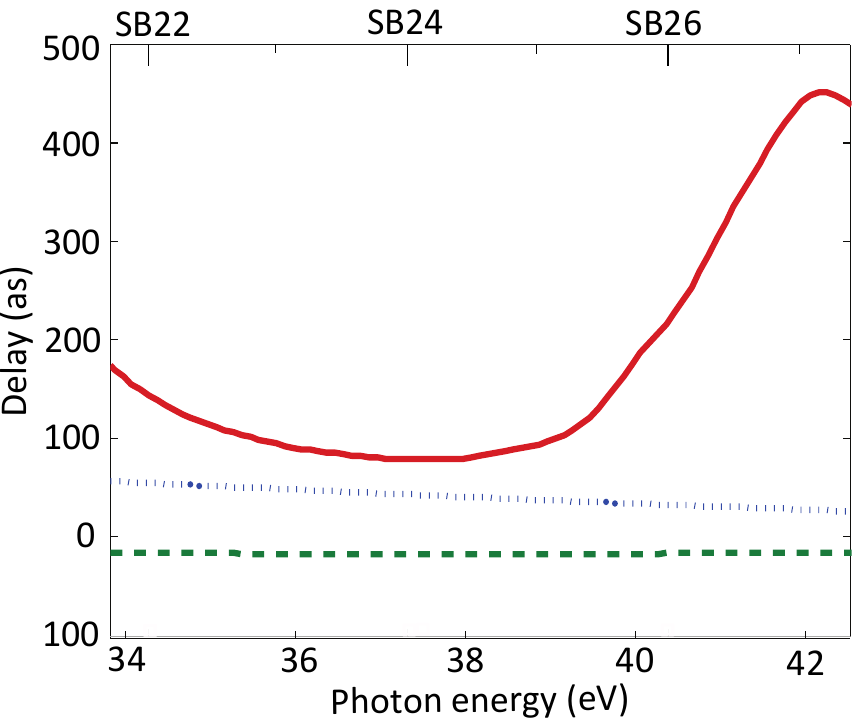}
  \caption{(color online) Ionization delay for the three angular channels $3p \to kd$ (blue dashed), $3p \to ks$ (green dashed) and $3s \to kp$ (red solid line).}
  \label{delay}
\end{figure}

Figure \ref{delay} presents calculated time delays $\tau^{(1)}$ for $3s\to kp$, $3p\to ks$ and $3p\to kd$ channels. The ionization delays from the $3p$ channel do not vary much with photon energy and remain small. The $3p \to ks$ delay is negligible while it takes about 70 as more time for the wave packet to escape towards the $d$ channel, due to the angular momentum barrier. The wave packet emitted from the $3s$ channel takes considerably more time to escape, especially in the region above 40 eV, owing to strong intershell correlation leading to screening by the $3p$ electrons.

\section{Comparison between theory and experiment}

\begin{figure}[htbp]
  \centering
  \includegraphics[width=8.0cm]{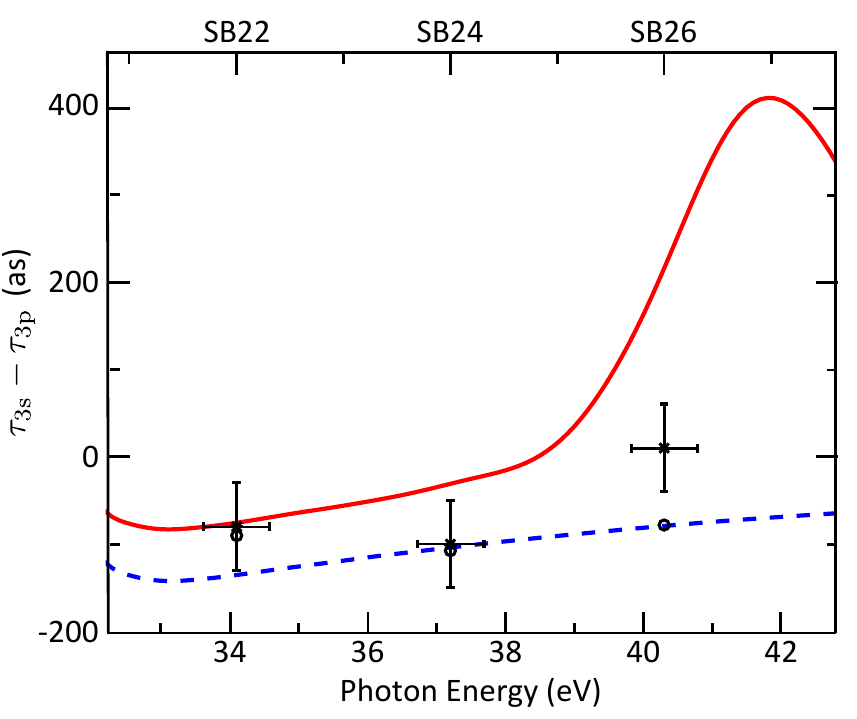}
\bigskip
\caption{(color online) Comparison between our theoretical calculations (dashed blue line, HF, red line, RPAE) and experiments (circles \cite{KlunderPRL2011}, crosses, present work)}
\label{comp}
\end{figure}

We present in Fig.~\ref{comp} a comparison between our experimental results (see Table 1) and our calculations.
The dashed blue and solid red lines refer to the independent-electron HF and RPAE calculations, respectively. The circles refer to the results of \cite{KlunderPRL2011}, while the other symbols (with error bars both in central energy and delay) are the results obtained in the present work.
Regarding the two sets of experimental results, they agree very well, except for that obtained at the highest energy corresponding to the sideband 26. Our interpretation is that we may be approaching the rapidly varying feature due to $3s-3p$ intershell correlation, leading to a large dispersion of the possible time delays (Note, however, that its position seems to be shifted compared to the RPAE result). The experimental and RPAE results agree well for the first sideband but less for the two higher energy sidebands. Surprisingly and perhaps accidentally the HF calculation gives there a closer agreement with the experiment.

We now discuss possible reasons for the discrepancy. Our calculation
of the influence of the dressing by the IR laser field is
approximate. It only uses the asymptotic form of the continuum
wave functions (both in the final and intermediate states), thereby
neglecting the effect of the core. This approximation should be tested
against theoretical calculations, and especially in a region where
correlation effects are important. We also neglect the influence of
the two-photon processes where the IR photon is absorbed (or emitted)
first \cite{VeniardPRA1996} [see Fig. \ref{Feynman} (e)]. The corresponding matrix elements are
usually small, except possibly close to a minimum of the cross
section, where the other process, usually dominant, is strongly
reduced. Interestingly, in such a scenario, the IR radiation would not simply be a probe used for the measurement
of the phase of a one-photon process, but would modify (control) the dynamics of the photoemission on an attosecond time scale.
Finally, in our theoretical calculation, correlation effects are only accounted for in the single ionization process (XUV absorption).
Additional correlation effects surrounding the probing, {\it e.g.} after the IR photon is absorbed might play a role.

In conclusion, the results shown above point out to
the need for explicit time-dependent calculations, which
would account for many-electron correlation and include not
only one-photon but also two-photon ionization. We also 
plan to repeat these experimental measurements
using attosecond pulses with a large and tunable bandwidth.
Our results demonstrate the potential of the experimental tools
using single attosecond pulses \cite{SchultzeScience2010} or
attosecond pulse trains \cite{KlunderPRL2011}. These
tools now enable one to measure atomic and molecular
transitions, more specifically, quantum phases and phase
variation, i.e. group delays, which could not be measured
previously.

\subsection*{Acknowledgements}
We thank G. Wendin for useful comments. This research was supported by the Marie Curie program ATTOFEL (ITN), the European Research Council (ALMA), the Joint Research Programme ALADIN of Laserlab-Europe II, the Swedish Foundation for Strategic Research, the Swedish Research Council, the Knut and Alice Wallenberg Foundation, the French ANR-09-BLAN-0031-01 ATTO-WAVE program, COST Action CM0702 (CUSPFEL).


\begin{thebibliography}{12}

\bibitem{CorkumNaturePhys2007}
P.~B. Corkum and F. Krausz, Nat. Phys.{\bf3}, 381 (2007).

\bibitem{KrauszRevModPhys2009}
F. Krausz and M. Ivanov, Rev. Mod. Phys. {\bf81}, 163 (2009).


\bibitem{SchmidtRPP1992}
V. Schmidt, Rep. Prog. Phys. {\bf 55}, 1483 (1992).

\bibitem{WignerPR1955}
E. P. Wigner, Phys. Rev. {\bf 98}, 145 (1955).

\bibitem{SchultzeScience2010}
 M. Schultze {\it et~al.}, Science {\bf 328}, 1658 (2010).

\bibitem{KheifetsPRL2010} A.S. Kheifets and I.A. Ivanov, Phys. Rev. Lett. {\bf 105}, 233002 (2010).

\bibitem{FanoPRA1985} U. Fano, Phys. Rev. A {\bf 32}, 617 (1985).

\bibitem{KlunderPRL2011}
K.Kl\"under {\it et~al.}, Phys. Rev. Lett. {\bf 106}, 143002 (2011).

\bibitem{Amusia1972361} M. Y. Amusia {\it et al.},
Phys. Lett. A {\bf 40}, 361 (1972).

\bibitem{MobusPRA1993} B. M\"obus {\it et al.}, Phys. Rev. A {\bf 47}, 3888 (1993).

\bibitem{GoulielmakisScience2004}
E.~Goulielmakis {\it et~al.}, Science {\bf 305}, 1267 (2004).

\bibitem{SansoneScience2006}
G.~Sansone {\it et~al.}, Science {\bf 314}, 443 (2006).

\bibitem{CavalieriNature2007}
Cavalieri, A. {\it et~al.}, Nature {\bf 449}, 1029 (2007).

\bibitem{PaulScience2001}
P.~M. Paul {\it et~al.}, Science {\bf 292}, 1689 (2001).

\bibitem{MairesseScience2003}
Y.~Mairesse {\it et~al.}, Science {\bf 302}, 1540 (2003).

\bibitem{YakovlevPRL2010} V.S. Yakovlev, J. Gagnon, N. Karpowicz and F. Krausz,
Phys. Rev. Lett. {\bf 105}, 073001 (2010).

\bibitem{NageleJPB2011} S. Nagele, R. Pazourek, J. Feist and J. Burgdorfer,
 J. Phys. B: Atom. Molec. Opt. Phys.  {\bf 44}, 081001 (2011); Phys. Rev. A {\bf 85}, 033401 (2012).

\bibitem{TaylorPRA2011} L. R. Moore, M. A. Lysaght,J. S. Parker, H. W. vanderHart and K. T. Taylor, Phys. Rev. A {\bf 84}, 061404 (2011).

\bibitem{ZhangPRA2011} C.-H. Zhang and U. Thumm, Phys. Rev. A {\bf 84}, 033401 (2011).

\bibitem{DahlstromPCCP2012} J. M. Dahlstrom {\it et al.}, J. Chem. Phys., in press (2012). \\ http://dx.doi.org/10.1016/j.chemphys.2012.01.017

\bibitem{IvanovPRL2011} M. Ivanov and O. Smirnova, Phys. Rev. Lett. {\bf 107}, 213605 (2011).

\bibitem{Kroon2012} D. Kroon {\it et al.}, in preparation.

\bibitem{Wendin1972} G. Wendin,  J. Phys. B: Atom. Molec. Opt. Phys.  {\bf 5}, 110 (1972).

\bibitem{LHuillierPRA1986} A. L'Huillier, L. J\"onsson and G. Wendin,  Phys. Rev. A {\bf 33}, 3938 (1986).

\bibitem{KennedyPRA1972} D.J. Kennedy and S.T. Manson, Phys. Rev. A {\bf 5}, 227 (1972).

\bibitem{KickasJES1996} A.Kikas {\it et~al.}, J. Electron Spectrosc. Rel. Phenom. {\bf 77} 241, (1996).

\bibitem{AC97} M.~Ia. Amusia and L.~V. Chernysheva,
 {\em Computation of atomic processes : {A} handbook for the {ATOM}
  programs}.
 Institute of Physics Pub., Bristol, UK, 1997.

\bibitem{CCR76} L.~V. Chernysheva, N.~A. Cherepkov, and V.~Radojevic,
Comp. Phys. Comm. {\bf 11}, 57 (1976).

\bibitem{CCR79}
L.~V. Chernysheva, N.~A. Cherepkov, and V.~Radojevic,
Comp. Phys. Comm. {\bf 18}, 87 (1979).

\bibitem{Samson2002265} J.A.R. Samson and W.C. Stolte, J. Electron Spectrosc. Rel. Phenom.
{\bf 123}, 265 (2002).


\bibitem{VeniardPRA1996}
V.~V\'eniard, R.~Ta\"ieb, and A.~Maquet,
Phys. Rev. A {\bf 54}, 721 (1996).

\end{thebibliography}


\end{document}